\documentclass[epjCONF,columns]{svjour} 
\usepackage{graphics}
\usepackage[varg]{txfonts} 
\usepackage[latin1]{inputenc}
\session-title{Hadron Collider Physics Symposium 2011}
\begin{document}
\title{CMS search for Standard Model Higgs: H$\to$W$^+$W$^-$ and H$\to$ZZ}
\author{Giuseppe Codispoti\inst{1}
  \fnmsep\thanks{\email{giuseppe.codispoti@cern.ch}} on behalf of the
  CMS Collaboration}
\institute{Universidad Aut\'onoma de Madrid, 28049 Madrid, Spain.}
\abstract{ A search for the Standard Model Higgs decaying to
  W$^+$W$^-$ and ZZ in pp collisions from LHC at $\sqrt{s}$ = 7 TeV
  using up to 1.7 fb$^{-1}$ of data recorded by the CMS detector is
  presented. This search covers a mass range from 110 GeV/c$^2$ to 600
  GeV/c$^2$. No significant excess above Standard Model background
  expectations is observed. Upper limits on the production cross
  section are derived.  }
%
\maketitle
\section{Introduction}
\label{intro}
The Standard Model (SM) is a very successful theory in describing
almost all phenomena in particle physics observed in past experiments.
One of the key remaining questions is the origin of the masses of
elementary particles attributed to the spontaneous breaking of
electroweak symmetry. The existence of the associated field quantum,
the Higgs boson (H), has still to be experimentally established. The
discovery or exclusion of the SM Higgs boson is one of the central
goals of the CERN Large Hadron Collider (LHC) physics program.

We report on the search for the SM Higgs decaying to W$^+$W$^-$ and ZZ
in pp collisions from LHC at $\sqrt{s}$ = 7 TeV using up to 1.7
fb$^{-1}$ of data recorded by Compact Muon Solenoid (CMS). A full
description of the detector as well as of the physics objects and
technical challenges to trigger interesting events and extract physics
signal from the high luminosity scenario can be found in \cite{CMS}.
%
%
The analysis covers a wide mass range hypothesis for the Higgs mass,
from 110 GeV/c$^2$ to 600 GeV/c$^2$.  The main challenge of the search
analysis is to distinguish the candidate signal from a wide set of
backgrounds showing the very same final states but much higher cross
sections. Beside to reducible backgrounds such as Z/W+jets, t$\bar{\rm
  t}$, single top, QCD hard scattering where jets are misidentified as
leptons, we have to deal also with irreducible backgrounds from WW and
ZZ production.
%
%
%
%
\section{Higgs decaying in W$^+$W$^-$ }
\label{sec:ww}
%
%
\begin{figure}
\resizebox{\columnwidth}{!}{%
  \includegraphics{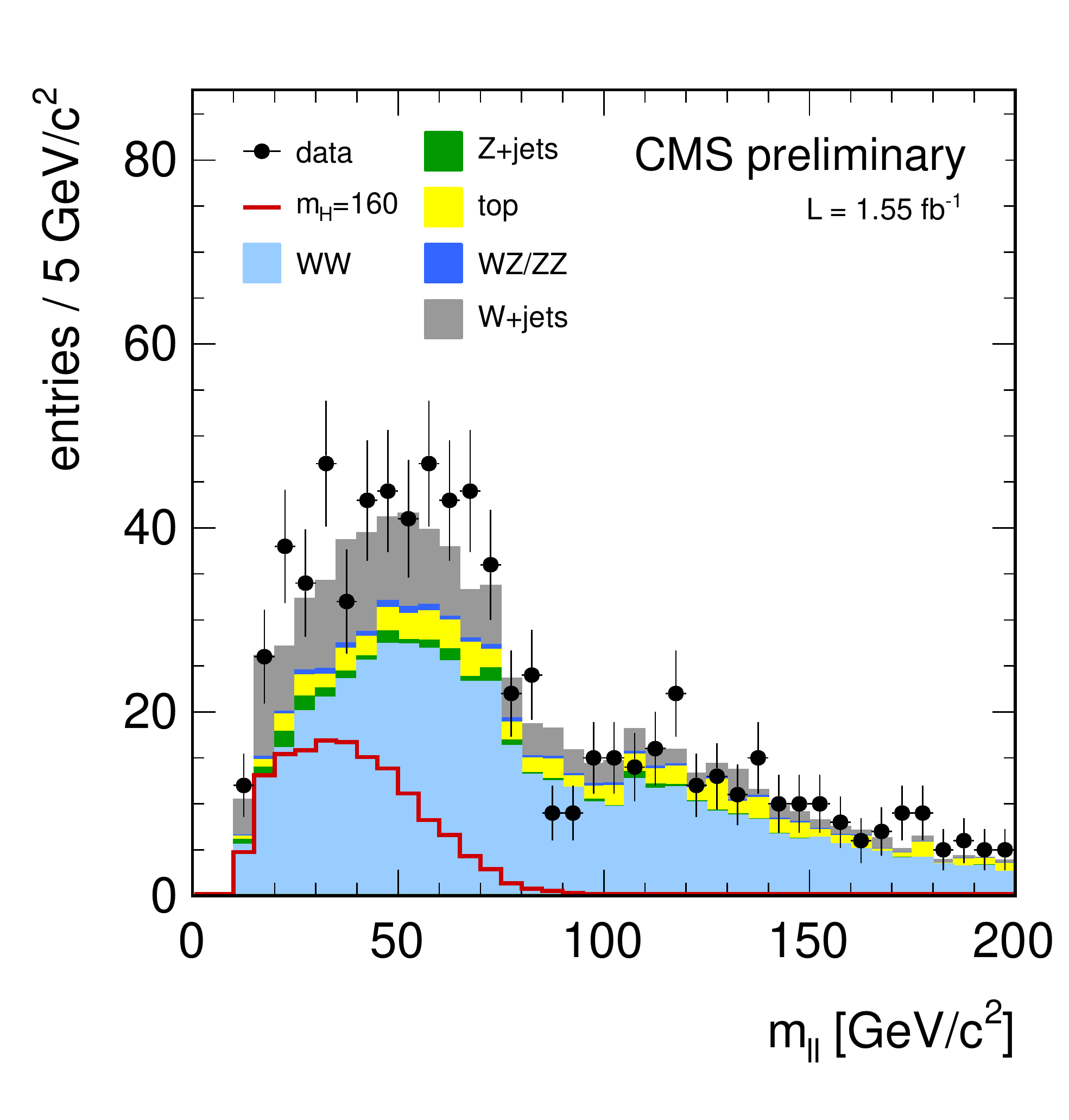} }
\caption{Di-lepton mass $m_{\rm ll}$ in the H $\to$ WW $\to$
  2$\ell$2$\nu$ search in the H+0 jet bin for an integrated luminosity
  of 1.55 fb$^{-1}$.}
\label{fig:ww_mll}
\end{figure}
\begin{figure}
\resizebox{\columnwidth}{!}{%
  \includegraphics{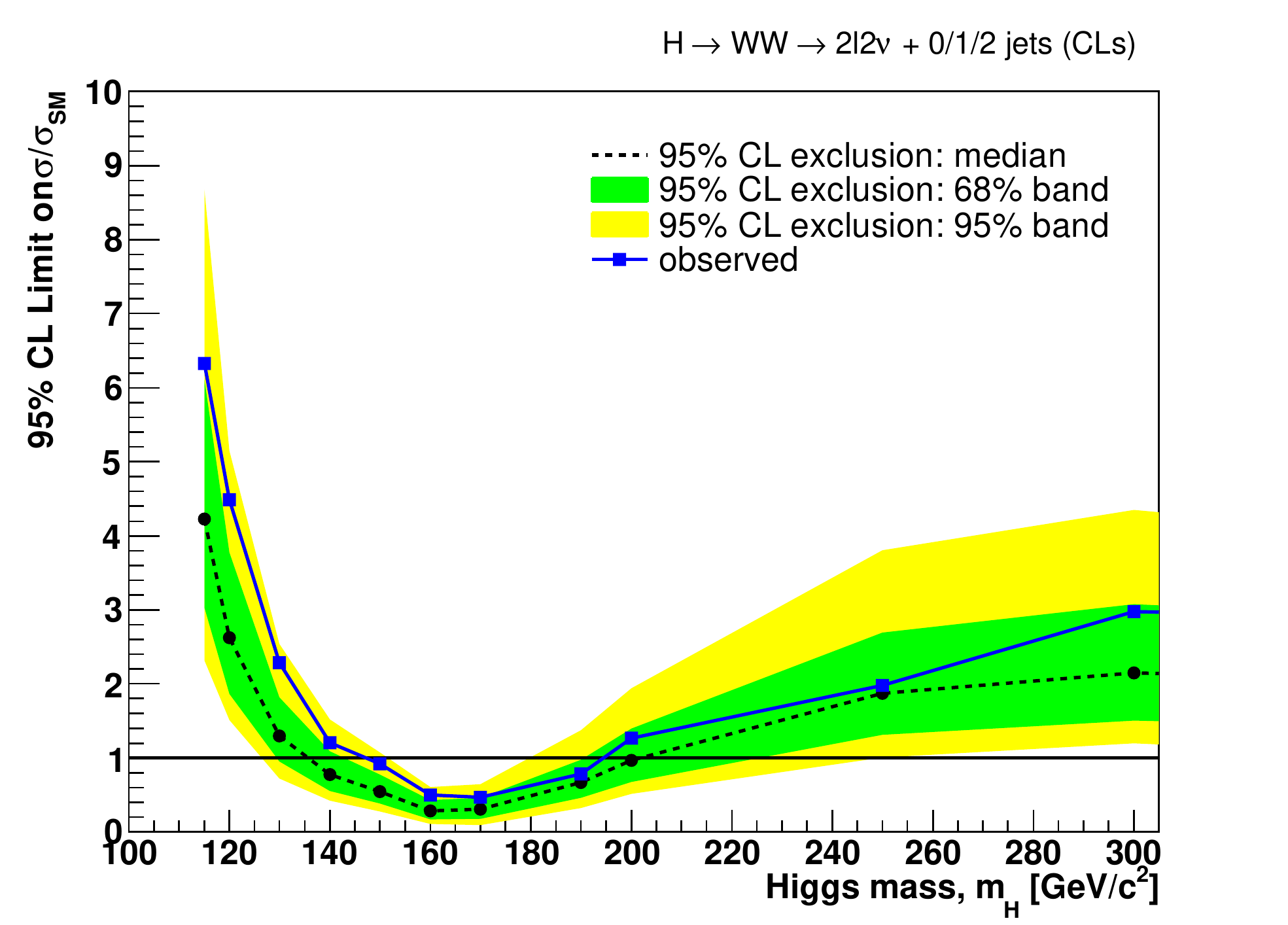} }
\caption{ The expected and observed upper limits at 95\% C.L. on
  \mbox{$\sigma$ (pp $\to$ H + X) $\times$ B(H $\to$ WW $\to$
    2$\ell$2$\nu$)} relative to the SM value for an integrated
  luminosity of 1.55 fb$^{-1}$ using the CLs approach.  The results
  are obtained using a cut-based event selection.}
\label{fig:HWW}
\end{figure}
The CMS search for \mbox{H $\to$ W$^+$W$^-$ \cite{H-WW}} selects
di-bosons candidates where both bosons decay leptonically, yielding an
experimental signature of two isolated, high transverse momentum
($p_{\rm T}$) oppositely-charged leptons (electrons or muons) and
large missing transverse energy ($E_{\rm T}^{\rm miss}$) due to
undetected neutrinos.  Given the large $E_{\rm T}^{\rm miss}$, a mass
peak cannot be reconstructed and therefore the analysis has a reduced
Higgs mass resolution.
The zero spin of the Higgs boson gives angular constraints to the
final state: the two leptons are expected with moderately small
opening angle while the $E_{\rm T}^{\rm miss}$ with large angle with
respect to the di-lepton system.
Due to different signal sensitivities, the events are separated into
three exclusive categories according to the jet multiplicity: H+0
jets, dominated by WW and W+jets backgrounds; H+1 jet, dominated by WW
an t$\bar{\rm t}$ background; and H+2 jets where the main background
is t$\bar{\rm t}$ and the main production mechanism is Vector Boson
Fusion.
The W$^+$W$^-$ non resonant contribution is reduced requiring a small
opening angle between the leptons. 
%
The remaining background is estimated using sidebands extrapolation
for the di-lepton mass distribution (fig. \ref{fig:ww_mll}) for Higgs
masses lower than 200 GeV/c$^2$ and Monte Carlo (MC) simulation for
the high masses, where less statistics is available.
The W+jets and QCD hard scattering background is controlled using a
control sample of loosely identified leptons extrapolated to the
signal region.  
Backgrounds induced by Z bosons are reduced requiring the di-lepton
mass to be outside the Z mass window. The remaining contribution is
estimated using the events inside the Z mass window, rescaled using
the ratio in/out estimated using simulation.
The t$\bar{\rm t}$ background is reduced rejecting events where a jet,
considering jets with $p_{\rm T} > 10$ GeV/c, can be identified as
b-quarks. The tagged sample, dominated by t$\bar{\rm t}$ and tW, is
used to extrapolate the residual contribution.
Results for 1.55 fb$^{-1}$ are shown in figure \ref{fig:HWW}: a SM
Higgs is excluded in the mass range from 147 GeV/c$^2$ to 194
GeV/c$^2$, a wide sensitivity on the full mass range is exploited in
the exclusion plot, even at low masses.
\section{Higgs decaying in ZZ }
\label{sec:zz}
The H $\to$ ZZ search involves the study of several final states.  The
general strategy identifies a first Z that decays into a pair
$\ell^+\ell^-$ and distinguish the case when the other decays into
$\ell^+\ell^-$ (4$\ell$), in two neutrinos (2$\ell$2$\nu$) or
hadronically (2$\ell$2q).  All the final states provide a strong
contribution at the high masses searches. Some of them may contribute
also to the low mass search, where one of the two bosons can be
off-shell. In particular the 4$\ell$ channel, despite the low yield,
provides very clean final states which can be then carefully inspected
for the Higgs signal search. In general a loss of sensitivity is
present around $m_{\rm H}$ = \mbox{180 GeV/c$^2$} due to the rapidly
decreasing branching ratio of \mbox{H $\to$ ZZ} in the SM.
\subsection{The four leptons final state}
\label{sec:4l}
\begin{figure}
\resizebox{\columnwidth}{!}{%
  \includegraphics{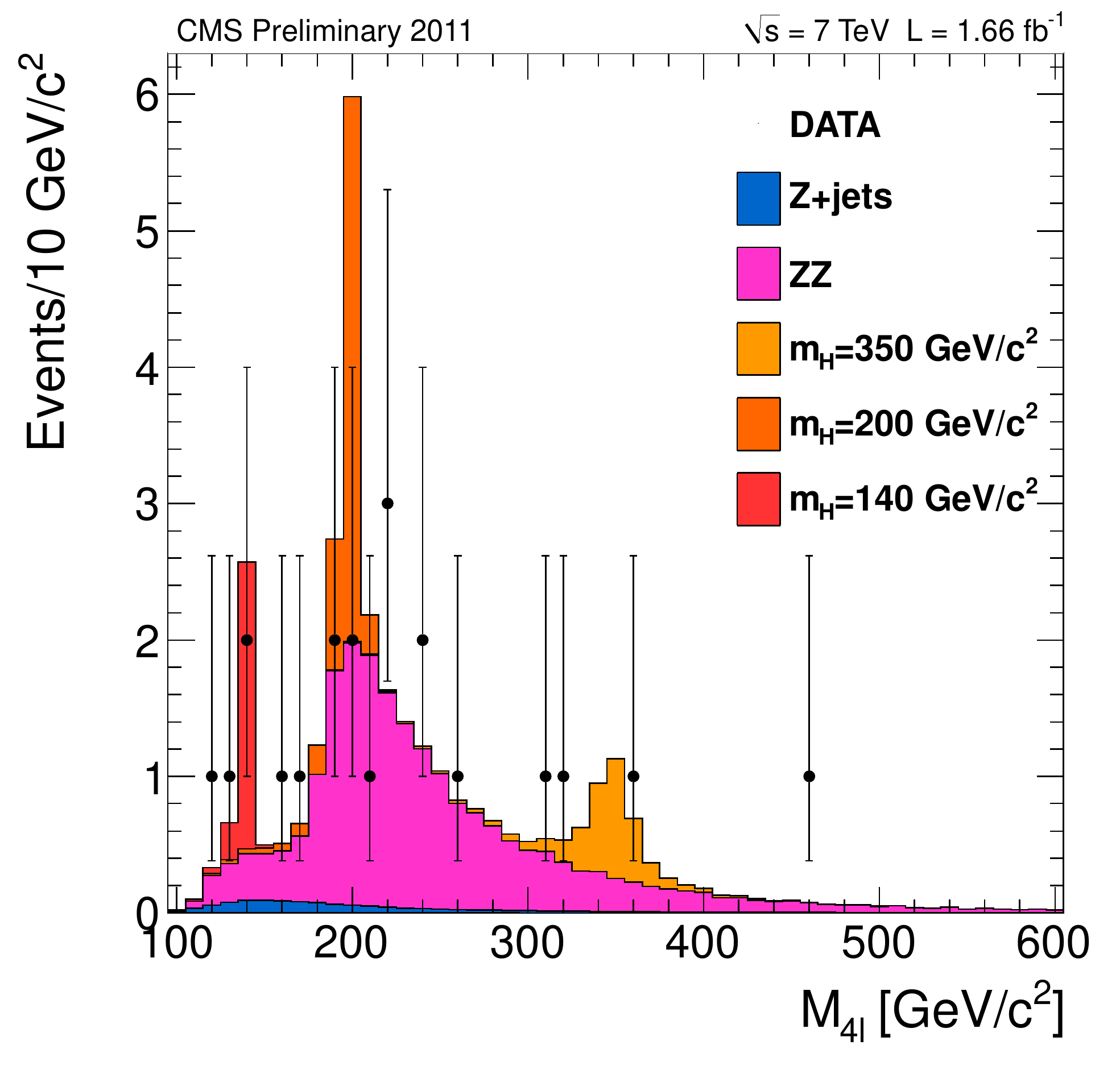} }
\caption{Distribution of the four-lepton reconstructed mass for the
  baseline selection in the sum of the 4$\ell$ channels. Points
  represent the data, shaded histograms represent the signal and
  background expectations. The samples correspond to an integrated
  luminosity of 1.66 fb$^{-1}$.}
\label{fig:4l_mass}
\end{figure}
\begin{figure}
\resizebox{\columnwidth}{!}{%
  \includegraphics{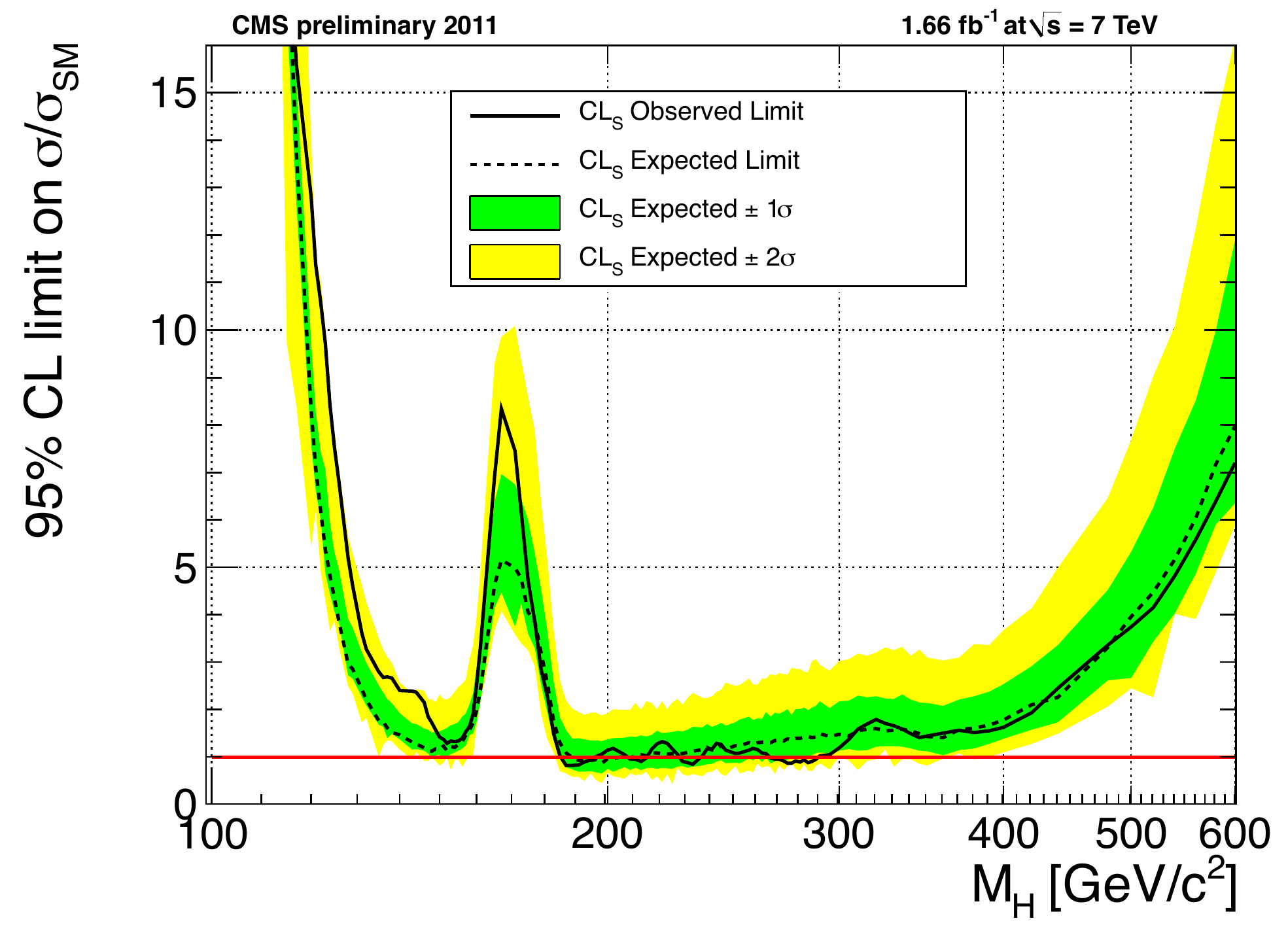} }
\caption{ The expected and observed upper limits at 95\% C.L. on
  \mbox{$\sigma$ (pp $\to$ H + X) $\times$ B(H $\to$ ZZ $\to$
    4$\ell$)} relative to the SM value for an integrated luminosity of
  1.66 fb$^{-1}$ using the CLs approach. The results are obtained
  using a shape analysis method.  }
\label{fig:4l}
\end{figure}
A clean signature of four well identified leptons belongs to the decay
channel \mbox{H $\to$ ZZ$^{(*)}$ $\to$ $\ell^\pm \ell^\mp \ell'^\pm
  \ell'^\mp$} with $\ell, \ell' = \rm e, \mu$ \cite{H-ZZ-4l}.  The
search relies solely on the measurement of leptons, and the analysis
achieves high lepton reconstruction, identification and isolation
efficiencies for a \mbox{ZZ$^{(*)}$ $\to$ 4$\ell$} system composed of
two pairs of same flavour and opposite charge isolated leptons,
e$^+$e$^-$ or $\mu^+\mu^-$.
On the other hand the small branching ratio associated to this decay
channel leads to a low events yield and requires a very careful study
of the background.
%
%
The first step of the analysis requires the identification of the best
Z candidate (Z$_1$) through the two opposite charge matching flavour
leptons with closest invariant mass to the nominal Z mass and
constrained to 60 GeV/c$^2$ $<$ $m_{\rm Z_1}$ $<$ 120 GeV/c$^2$. Then
at least two other leptons are required for the second Z
(Z$_2$). Isolation and impact parameter requests, imposing the leptons
to be originated from the same primary vertex further reduce the
background.  A baseline selection, for the analysis in the full mass
range imposes 20 GeV/c$^2$ $<$ $m_{\rm Z_2}$ $<$ 120 GeV/c$^2$. An
high-mass selection, which is used for $m_{\rm H} > 2 \times$ $m_{\rm
  Z}$ and the ZZ cross section measurement, requires 60 GeV/c$^2$ $<$
$m_{\rm Z_2}$ $<$ 120 GeV/c$^2$.
The ZZ background estimation is performed using the MC prediction at
Next to Leading Order (NLO) that gives a good description of the mass
distribution shape. The MC prediction is then normalized using
\mbox{Z$ \to \ell^+\ell^-$} events in data and the expected ratio
$\sigma_{\rm ZZ}/\sigma_{\rm Z}$ and acceptance obtained from MC.
The fake leptons from Z+jets instrumental background are estimated
using a control sample containing the Z$_1$ plus loosely identified
leptons. The Zb$\bar{\rm b}$ and Zt$\bar{\rm t}$ background are
estimated removing flavour, charge and isolation request for Z$_2$ and
reversing the impact parameter cut.
The invariant mass distribution of the events surviving the selection
is shown in figure \ref{fig:4l_mass}.  Results of the analysis are
summarized in the exclusion plot in figure \ref{fig:4l}.
\subsubsection{The two leptons two taus final state}
\begin{figure}
\resizebox{\columnwidth}{!}{%
  \includegraphics{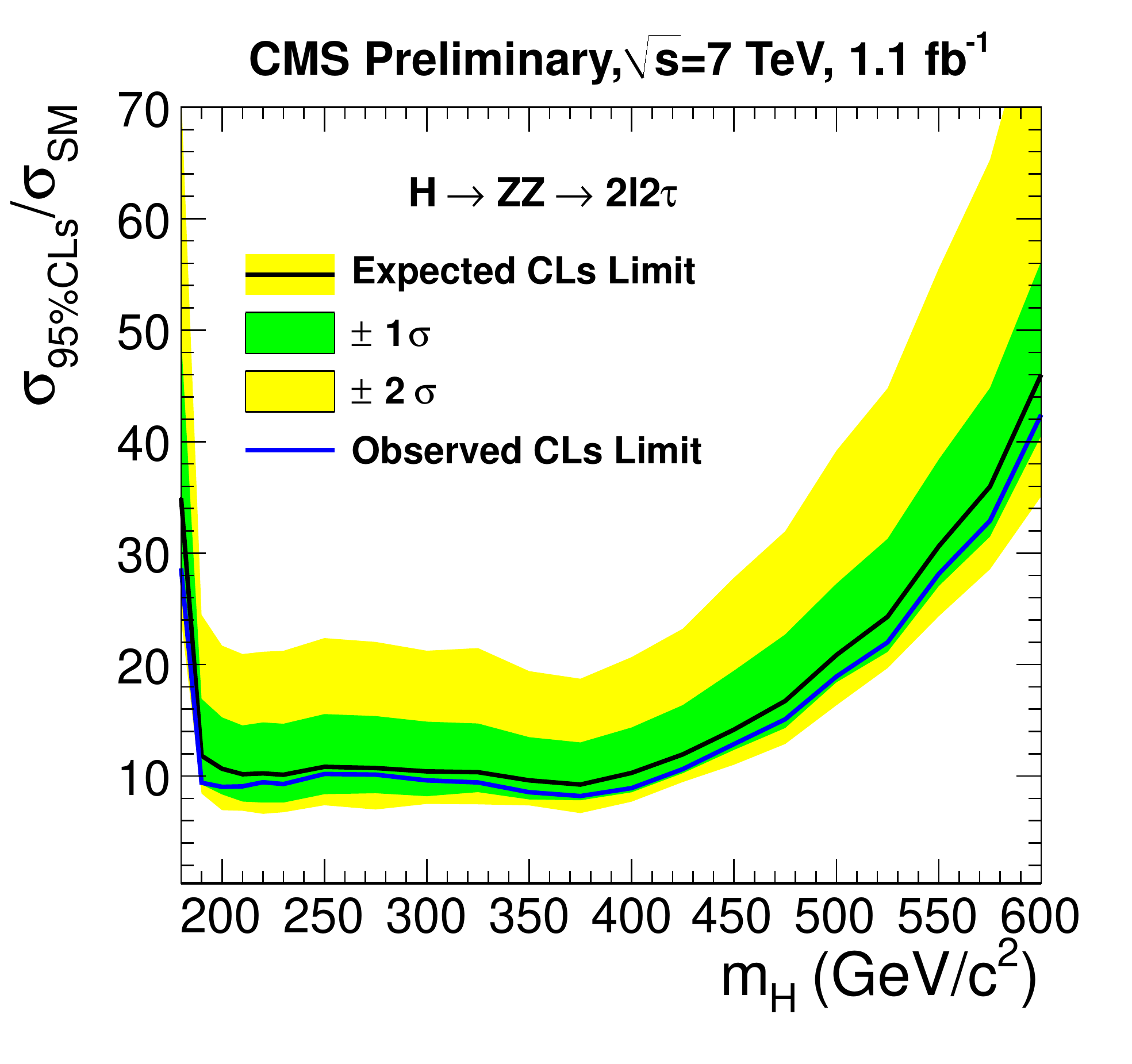} }
\caption{ The expected and observed upper limits at 95\% C.L. on
  \mbox{$\sigma$ (pp $\to$ H + X) $\times$ B(H $\to$ ZZ $\to$
    2$\ell$2$\tau$)} relative to the SM value for an integrated
  luminosity of 1.1 fb$^{-1}$ using the CLs approach.
}
\label{fig:2l2tau}
\end{figure}

A specific analysis completes the four leptons scenario with the
specific case where the secon Z decays in two $\tau$ leptons
\cite{H-ZZ-2l2tau}.  The analysis is performed on a data sample where
4$\ell$ events are already discarded. Additional requirements are
added in order to be able to reconstruct the 2 $\tau$.  Resuls are
shown in figure \ref{fig:2l2tau}.
\subsection{The two leptons two neutrinos final state}
\label{sec:2l2nu}
\begin{figure}
\resizebox{\columnwidth}{!}{%
  \includegraphics{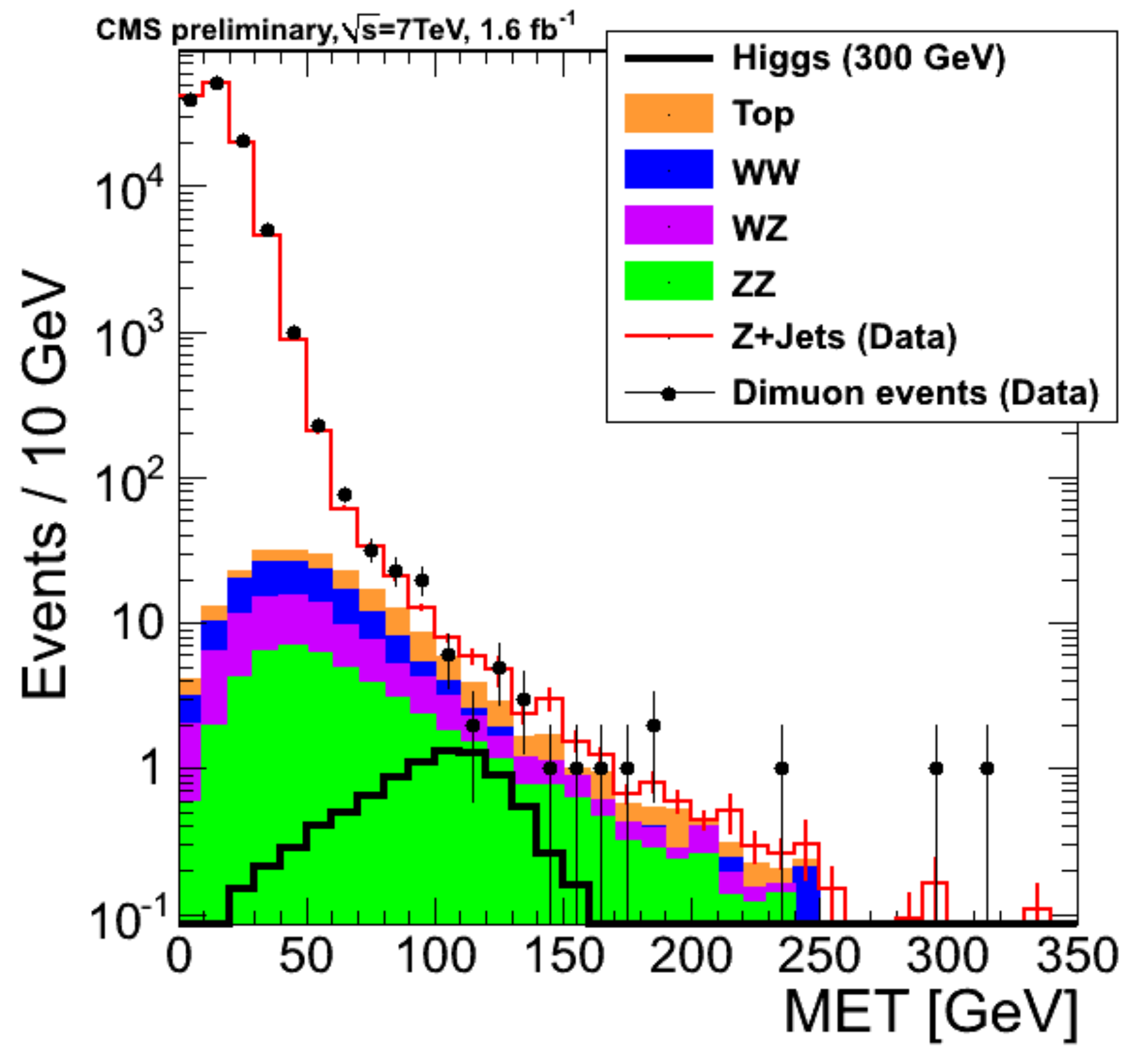} }
\caption{Missing transverse energy distribution in the \mbox{H $\to$
    ZZ $\to$ 2$\mu$2$\nu$} channel at pre-selection level for 1.6
  fb$^{-1}$ of data. }
\label{fig:2l2nu_met_mu}
\end{figure}

\begin{figure}
\resizebox{\columnwidth}{!}{%
  \includegraphics{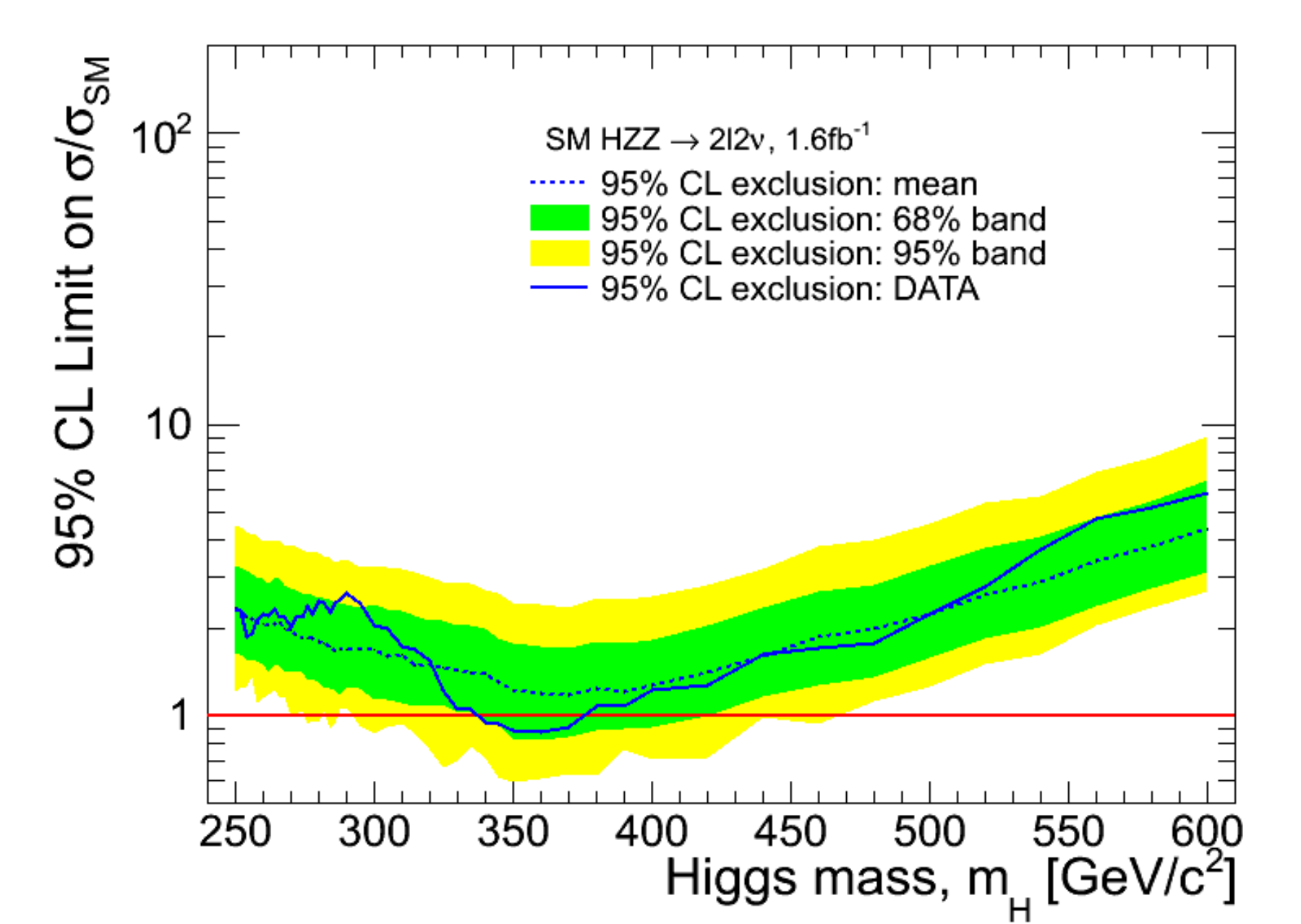} }
\caption{ The expected and observed upper limits at 95\% C.L. on
  \mbox{$\sigma$ (pp $\to$ H + X) $\times$ B(H $\to$ ZZ $\to$
    2$\ell$2$\nu$)} relative to the SM value for an integrated
  luminosity of 1.6 fb$^{-1}$ using the CLs approach.  The results are
  obtained using a cut-based event selection.}
\label{fig:2l2nu}
\end{figure}

High branching ratio and good sensibility to the high mass range
(250-600 GeV/c$^2$) characterize the decay channel \mbox{H $\to$
  ZZ$^{(*)} \to \ell^+ \ell^- \nu\bar{\nu}$} \cite{H-ZZ-2l2nu} where
the second Z decays in two neutrinos.  Two well isolated, close angle,
charged leptons come from the decay of the first boosted Z. The second
Z decaying in two neutrinos brings to large $E_{\rm T}^{\rm miss}$,
which characterizes the event and provides a good background
suppression power as shown in figure \ref{fig:2l2nu_met_mu}.
The t$\bar{\rm t}$ background is suppressed rejecting b-tagged
events. The WZ and ZZ backgrounds are estimated from MC
simulation. Residual backgrounds are estimated from data: Z+jets is
modeled using photon + jets events and t$\bar{\rm t}$ and WW are
estimated events with different flavour leptons (e$\mu$ events).
Cuts optimization is performed on the basis of the Higgs mass
hypothesis. The dilepton pair is constrained to the Z mass, the
$E_{\rm T}^{\rm miss}$ is requested to be not aligned with jets in
order to cope with jet energy mis-measurements.  Transverse mass is
also used to characterize the sample.
Results of this analysis are shown in figure \ref{fig:2l2nu}.
\subsection{The two leptons two jets final state}
\label{sec:2l2q}
\begin{figure}
\resizebox{\columnwidth}{!}{%
  \includegraphics{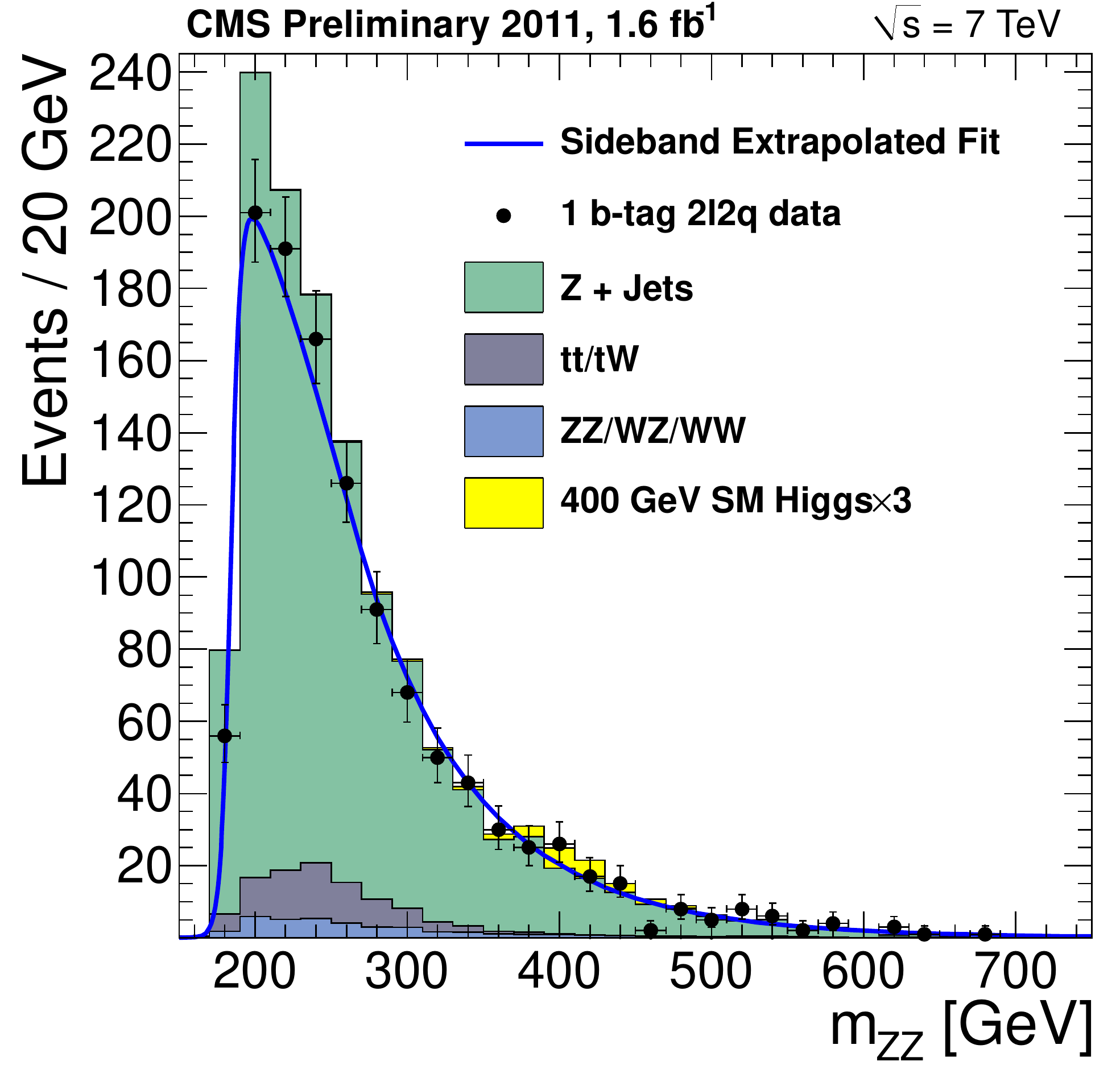} }
\caption{The $m_{ZZ}$ invariant mass distribution after final
  selection in the 1 b-tag category of the H $\to$ ZZ$^{(*)}$ $\to$
  2$\ell$2q channel.  Points with error bars show distributions of
  data, solid histograms show expectation from simulated events, solid
  curved line shows prediction of background from sideband
  extrapolation procedure. The samples correspond to an integrated
  luminosity of 1.6 fb$^{-1}$. }
\label{fig:2l2q_mzz_1btag}
\end{figure}
\begin{figure}
\resizebox{\columnwidth}{!}{%
  \includegraphics{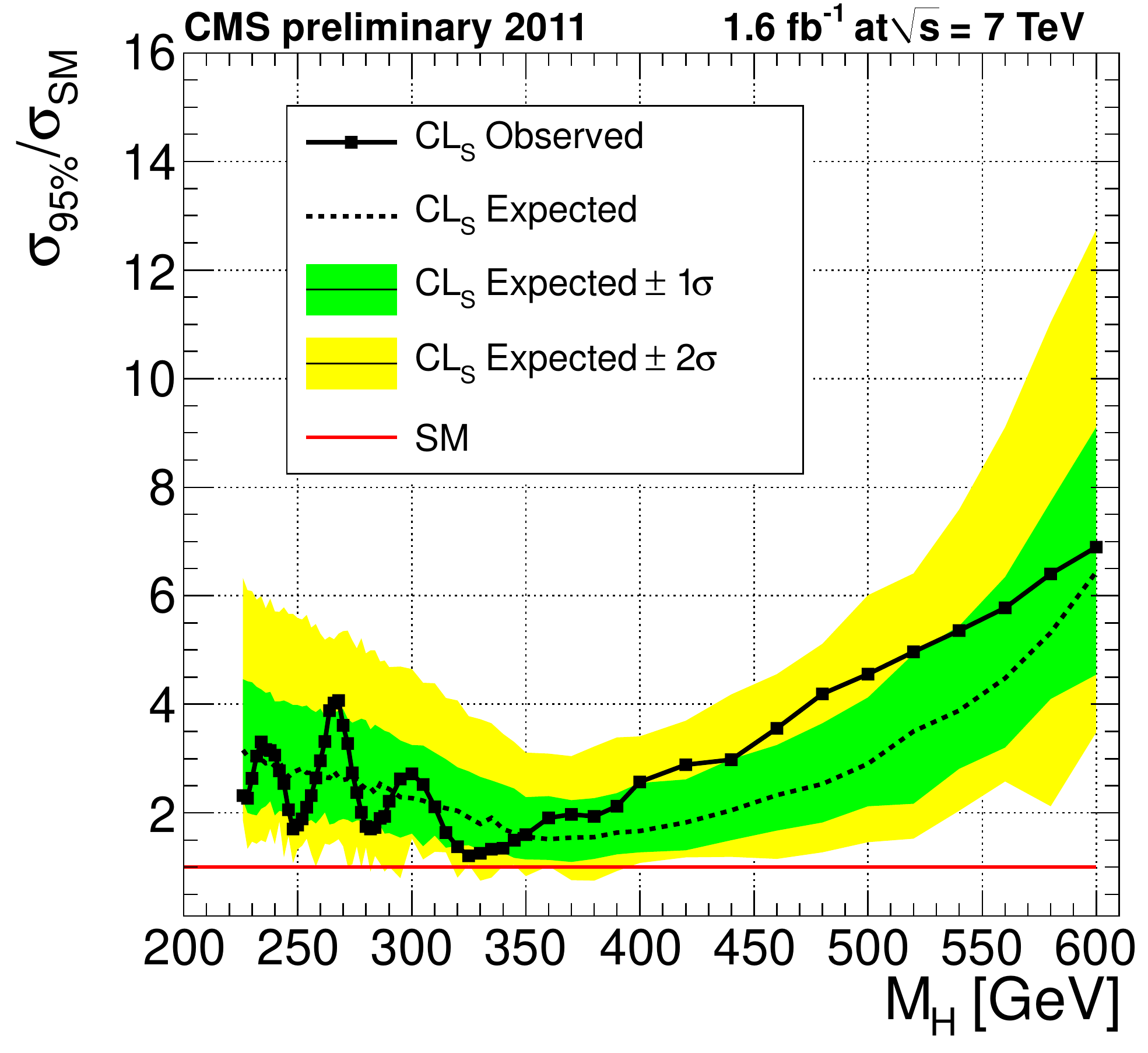} }
\caption{ The expected and observed upper limits at 95\% C.L. on
  \mbox{$\sigma$ (pp $\to$ H + X) $\times$ B(H $\to$ WW $\to$
    2$\ell$2q)} relative to the SM value for an integrated luminosity
  of 1.6 fb$^{-1}$ using the CLs approach.  The results are obtained
  using a shape analysis method. }
\label{fig:2l2q}
\end{figure}
The advantage of the channel where one Z decays hadronically is the
fully reconstructable final state with two leptons and two jets
\cite{H-ZZ-2l2q}.  Moreover, the branching fraction of the decay
channel \mbox{H $\to$ ZZ$^{(*)}$ $\to \ell^+ \ell^-$ q$\bar{\rm q}$}
is about 20 times higher than the 4$\ell$ channel. This may lead to
better sensitivity to SM Higgs boson production at high masses, where
kinematic requirements can effectively suppress background.
For the analysis we select events containing at least two leptons and
two jets compatible with the Z mass hypothesis.  Since the final
states are fully reconstructed we can perform angular analysis: the
initial zero spin of the Higgs bosons constraint the relative angles
between the decays product. A dedicated likelihood is used to
discriminate signal compatible events.
Since Z+jets events involve gluon radiation we can use gluon-jets
characteristics to reject them. In order to do this we build a
likelihood discriminant based on number of neutral and charged tracks
and their transverse momentum distribution.  The effectiveness of the
gluon-quark likelihood discriminator is tested on photon+jets sample
where jets mainly come from quarks.
The t$\bar{\rm t}$ background is suppressed requesting small $E_{\rm
  T}^{\rm miss}$ and controlled using the e$\mu$ sample.
Finally the presence of b-jets leads to a better characterization of
the background. For this reason we divide the data sample in three
main categories: \mbox{0 b-tag} category, were events do not contain
jets satisfying the b-tag algorithm; we use the quark-gluon likelihood
discriminator to veto gluons events which are studies in a separate
category; 1 b-tag category; 2 b-tag category.
The background determination is completely data-driven: we use
sidebands extrapolation from di-jet mass distribution, where sidebands
are defined as \mbox{(60 GeV/c$^2$} $< m_{jj} < $ \mbox{75 GeV/c$^2$)}
$\cup$ \mbox{(105 GeV/c$^2$} $< m_{jj} < $ \mbox{130 GeV/c$^2$)}. This
method better describe data than the MC simulation based as can be
easily seen in figure \ref{fig:2l2q_mzz_1btag}.
Results of this analysis are shown in figure \ref{fig:2l2q}.
\section{Conclusions}
\label{sec:concl}
\begin{figure}
\resizebox{\columnwidth}{!}{%
  \includegraphics{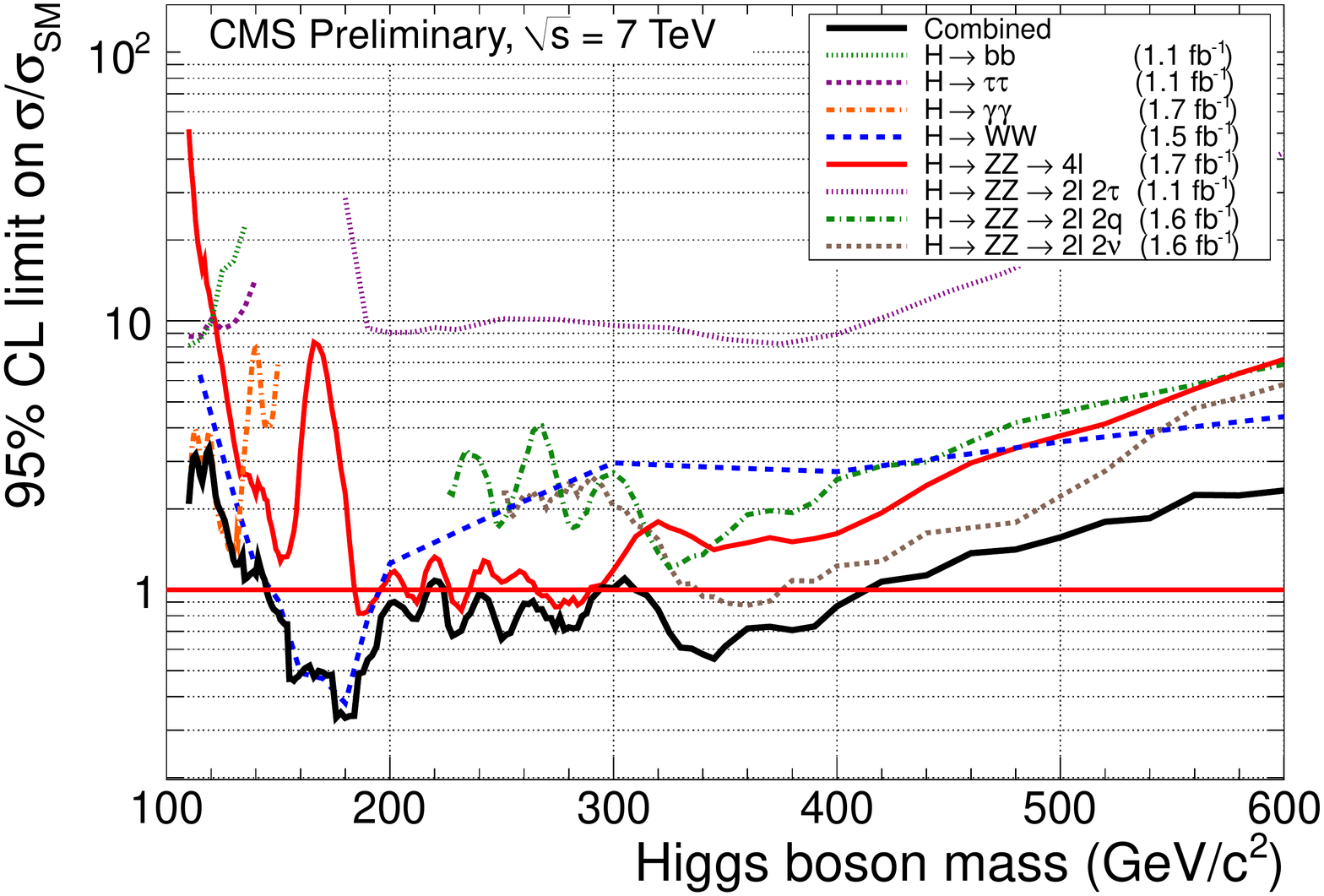} }
\caption{The observed 95\% C.L. upper limits on the signal strength
  modifier $\mu = \sigma/\sigma_{\rm SM}$ as a function of the SM
  Higgs boson mass for the eight CMS major analyses and their
  combination, included the five described in this paper.  The limits
  are obtained with the CLs method.}
\label{fig:comb}
\end{figure}

A search for the SM Higgs boson decaying into two Z or W bosons has
been presented.  The analisys is performed with the first 1.6
fb$^{-1}$ of data recorded by CMS in 2011 over a total of almost 5
fb$^{-1}$ and cover a mass range from \mbox{130 GeV/c$^2$} to
\mbox{440 GeV/c$^2$}. No significant excess is observed thus no
evidence of a Higgs boson is found. Combining the results from the
different channels \cite{H-WW-ZZ-comb} we set limits on the Standard
Model Higgs production in the ranges \mbox{145-216 GeV/c$^2$},
\mbox{226-288 GeV/c$^2$} and 310-400 GeV/c$^2$. The limits and the
contribution of the decay channels illustrated are summarized in
figure \ref{fig:comb}.
\end{document}